\documentclass{article}
\usepackage{ijcai26}

\usepackage{times} 
\usepackage{soul}  
\usepackage{url} 
\usepackage[hidelinks]{hyperref} 
\usepackage[utf8]{inputenc}
\usepackage[small]{caption} 
\usepackage{graphicx}
\usepackage{amsmath} 
\usepackage{amsthm}
\usepackage{booktabs}
\usepackage{algorithm}
\usepackage{algorithmic}
\usepackage[switch]{lineno}
\usepackage{amssymb}
\usepackage{amsfonts}
\usepackage{multirow}
\usepackage{tabularx}
\usepackage{enumitem}

\title{The Dawn of Agentic EDA: A Survey of Autonomous Digital Chip Design}

\author{
Zelin Zang$^{1,3}$
\and
Yuhang Song$^2$
\and
Aili Wang$^2$
\and 
Bingo Wing-Kuen Ling$^1$
\and
Qi Sun$^4$
\and
Zhen Lei$^5$ 
\and
Fuji Yang$^1$
\and
Cheng Zhuo$^4$  
\and
Jiebo Luo$^5$  
\affiliations 
$^1$Center for the Integrated Circuits and Artificial Intelligence, Tsientang Institute for Advanced Study\\
$^2$ZJUI Institute, Zhejiang University\\
$^3$AI Lab, Westlake University \\
$^4$College of Integrated Circuit, Zhejiang University \\
$^5$Hong Kong Institute of Science \& Innovation \\
} 
 
\begin{document}  
 
\maketitle

\begin{abstract}  
The semiconductor industry faces a critical ``\mbox{Productivity} Gap'' where design complexity outpaces human capacity. While the ``AI for EDA'' revolution (L2) successfully optimized specific point problems, a paradigm shift toward \textbf{Agentic EDA} (L3) is emerging, evolving from passive prediction to autonomous orchestration of the RTL-to-GDSII flow. This survey presents the first systematic framework for this transition, framing Agentic EDA not merely as ``Chat with Tools,'' but as a \textbf{Constrained Neuro-Symbolic Optimization} problem. We propose a novel taxonomy rooted in a Cognitive Stack—\textit{Perception} (aligning multimodal semantics), \textit{Cognition} (planning under strict constraints), and \textit{Action} (deterministic tool execution)—to dissect how probabilistic agents navigate zero-tolerance physical laws. Through this lens, we analyze the landscape: (1) in Frontend, the shift from one-shot generation to dual-loop syntactic-semantic repair; (2) in Backend, the dichotomy between algorithm-centric solvers and agent-centric orchestrators that treat executable code as a latent space. Finally, we critically examine the Trustworthiness gap, advocating for Sim-to-Silicon benchmarks and formal grounding to transform brittle prototypes into resilient engineering systems.
\end{abstract}  

\section{Introduction: The Inevitable Rise of Agentic EDA}
\label{sec:intro}

The semiconductor industry is approaching a critical threshold where the complexity of System-on-Chip (SoC) design threatens to outpace human capability. This ``Design Productivity Gap,'' first identified decades ago, has widened as transistor counts surpass hundreds of billions while design teams grow only linearly \cite{markov_limits_2014}. Historically, the Electronic Design Automation (EDA) community has responded with waves of abstraction: from manual layout (L0) to algorithmic automation (L1) pioneered by Simulated Annealing \cite{kirkpatrick_optimization_1983} and logic synthesis \cite{brayton_logic_1984}. In the past decade, the ``ML for EDA'' revolution (L2) introduced data-driven prediction, exemplified by deep learning for placement \cite{lin_dreamplace_2019,lai_maskplace_2022,mirhoseini_graph_2021} and congestion analysis \cite{xie_routenet_2018}. However, these Copilot-level tools \cite{liu_chipnemo_2023} typically function as \textit{Specialized Solvers}—they optimize specific metrics (e.g., wirelength) within a siloed stage, lacking the global context to resolve cross-stage conflicts such as how RTL logic changes affect routing congestion.

Concurrently, Software Engineering (SWE) has undergone a tectonic shift: LLMs have evolved from passive code completion to autonomous software engineers. State-of-the-art agents like SWE-agent \cite{yang2024sweagent} can now explore file systems, write reproduction scripts, and submit pull requests, successfully resolving real-world GitHub issues in popular libraries (e.g., scikit-learn, Django) on the SWE-bench leaderboard \cite{jimenez2024swebench}. This transition from local syntax suggestion to global semantic reasoning inspires the next leap in hardware design.

We argue that the industry is now witnessing a necessary transition to \textbf{Agentic EDA} (L3+), driven not merely by the availability of Large Language Models (LLMs), but by the fundamental limitations of siloed optimization. Unlike games where rules are explicit, chip design is a constrained neuro-symbolic problem where ``correctness'' is binary (e.g., Timing Closure) yet the search space is non-convex and larger than Go \cite{mirhoseini_graph_2021}.

\begin{figure*}[t]
    \centering
    \includegraphics[width=0.999\linewidth]{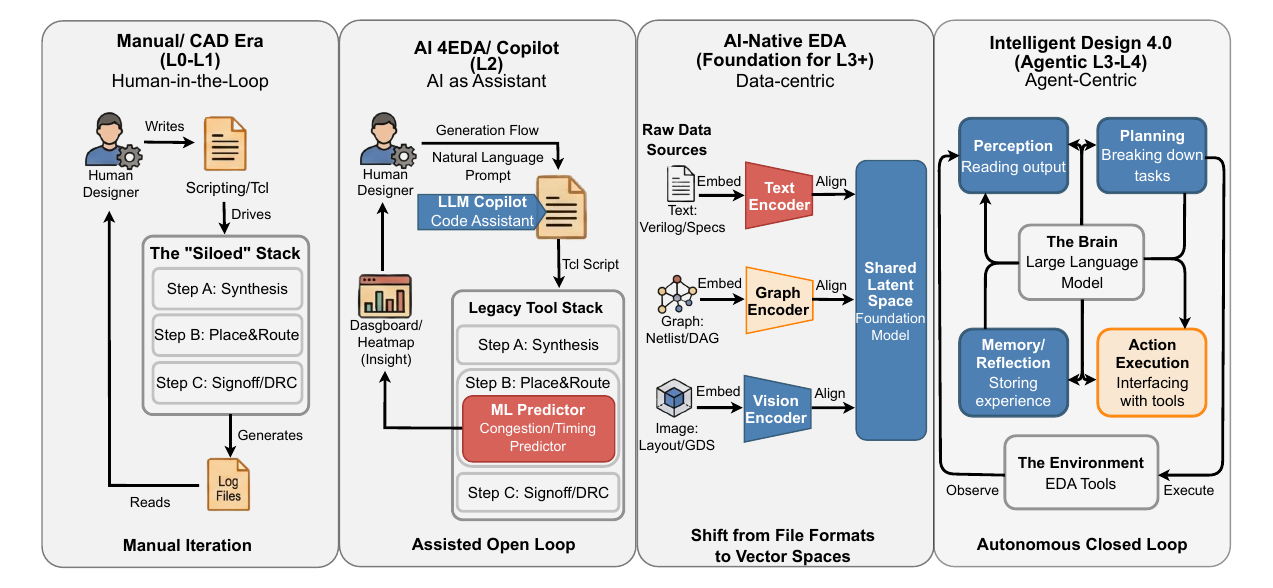} 
    \vspace{-0.8cm}
    \caption{\textbf{Evolution of IC design paradigms.} We delineate the trajectory from (1) \textbf{Manual/CAD (L0-L1)} to (2) \textbf{AI Copilots (L2)} which serve as assistants, and finally to (3) \textbf{Agentic EDA (L3-L4)} which achieves autonomy. This shift moves from \textit{optimizing point tools} to \textit{automating workflows} by incorporating reasoning and planning. A key enabler is the transition from pixel-based GUIs to code-based orchestration (Tcl/Python), treating EDA tools as executable environments.}
    \label{fig:evolution}
\end{figure*}

\paragraph{The Evolution: From Prediction to Orchestration.}
To clarify this paradigm shift, we propose a taxonomy of EDA autonomy levels, analogous to SAE levels in autonomous driving \cite{sae_j3016_2021,yurtsever_survey_2020} (Figure \ref{fig:evolution}):

\begin{description}[leftmargin=*, style=unboxed]
    \item[L0-L1: Tool-Assisted Human Design (Manual \& Algorithmic).] 
    \textit{Human defines flow, Tool solves sub-problems.} L0 represents purely manual layout (pre-1980s). L1 introduced algorithmic solvers (e.g., logic synthesis, placement) where humans provide constraints and tools perform mathematical optimization \cite{brayton_logic_1984}.
    
    \item[L2: AI-Assisted Copilots (Prediction).]
    \textit{Human drives flow, AI predicts outcomes.} This era (2018-2023) integrates ML to accelerate specific steps (e.g., congestion prediction \cite{xie_routenet_2018}) or tune parameters (e.g., Bayesian Optimization). The AI acts as a consultant, offering suggestions without execution authority.

    \item[L3: Agentic EDA (Orchestration).]
    \textit{AI drives tasks, Human monitors.} The current frontier where agents autonomously decompose tasks, generate scripts (RTL/Tcl), and interact with EDA tools to close loops (e.g., fixing syntax errors or hold violations) \cite{lu_rtllm_2024}. The agent possesses ``hands'' (tool-use) and ``eyes'' (log reading).

    \item[L4: Autonomous Design 4.0 (Self-Driving).]
    \textit{AI drives flow, Human defines intent.} The future state where systems autonomously navigate the full RTL-to-GDSII flow, treating the entire toolchain as a differentiable or black-box environment to optimize for PPA (Power, Performance, Area) with zero human intervention \cite{autochip}.
\end{description}

\paragraph{Challenges: The ``Code-Reality Gap''.}
Adapting LLMs to this domain faces a unique ontology mismatch. LLMs are trained on open-ended text, whereas chips are governed by unforgiving physical laws (Kirchhoff's laws) and zero-tolerance constraints.
\begin{itemize}[leftmargin=*, noitemsep, topsep=2pt]
    \item \textit{Semantic Blindness}: A model can generate syntactically correct Verilog that is functionally useless due to timing violations.
    \item \textit{Data Proprietary}: High-quality industrial data (e.g., 5nm tape-out logs) is siloed, unlike the open web.
    \item \textit{The Reality Gap}: While RL-based systems (e.g., AlphaChip) have achieved industrial tape-out, current LLM-based autonomous agents have yet to demonstrate a complete, zero-human-in-the-loop tape-out of an industrial-grade chip.
\end{itemize} 
Thus, Agentic EDA is not just ``Chat with PDF'' but a system of constrained optimization requiring rigorous grounding \cite{he_large_2025}.

\paragraph{Scope \& Contributions.}
While recent surveys have covered Circuit Foundation Models \cite{fang_survey_2025} or general LLM-EDA applications \cite{pan_survey_2025}, this paper specifically targets the \textit{Agentic Execution Layer}—focusing on \textit{how} to build autonomous workflows rather than just \textit{what} models can generate. We trace the lineage from classical algorithms to modern agents, providing a novel taxonomy that segments advancements into {Neural-Augmented Solvers} and {Agentic Orchestrators}. This includes a systematic review analyzing over 40 representative frameworks from code generation \cite{liu_verilogeval_2023} to physical design \cite{mirhoseini_graph_2021}. Finally, we propose a future roadmap to critically analyze the ``Sim-to-Silicon Gap,'' arguing that lack of industrial-grade benchmarks is the primary barrier to adoption. To navigate this landscape, we present a ``Survey Map'' in Table \ref{tab:eda_overview_map}.

\begin{table*}[t]
    \centering
    \caption{Taxonomy-aligned overview of representative agentic EDA works. \textbf{Type} corresponds to the classification in Sec.~\protect\ref{sec:foundations}: \textbf{PRT} = Prompt-Based Reasoning; \textbf{SFT} = Fine-Tuned Specialization; \textbf{MAS} = Multi-Agent Orchestration; \textbf{FDN} = Circuit Foundation Model. The table further characterizes each work by its functional \textbf{Role}, input-output \textbf{Info Flow}, and algorithmic \textbf{Method Form}, highlighting the diverse integration strategies of LLMs in chip design.}
    \vspace{-0.3cm}
    \label{tab:eda_overview_map}
    \scriptsize
    \setlength{\tabcolsep}{1.5pt}
    \resizebox{\textwidth}{!}{
    \begin{tabular}{@{}p{0.20cm}>{\raggedright\arraybackslash}p{3.8cm}p{0.6cm}p{1.2cm}p{2.0cm}p{1.9cm}>{\raggedright\arraybackslash}p{2.3cm}p{1.9cm}@{}}
        \toprule
        \multirow{2}{*}{\rotatebox[origin=c]{90}{\textbf{Stage}}} & \multirow{2}{*}{\textbf{Paper}}             & \multirow{2}{*}{\textbf{Type}} & \multirow{2}{*}{\textbf{Role}} & \multicolumn{2}{c}{\textbf{Info Flow}} & \multirow{2}{*}{\textbf{Method Form}} & \multirow{2}{*}{\textbf{Eval (typ.)}}                      \\
        \cmidrule(lr){5-6}
                                                                  &                                             &                                &                                & \textbf{Input}                         & \textbf{Output}                       &                                       &                    \\
        \midrule
        \multirow{9}{*}{\rotatebox[origin=c]{90}{Frontend}}
                                                                  & Spec2RTL-Agent \cite{yu_spec2rtl_2025}      & PRT                            & Agent                          & PDF Spec                               & Verilog + Plan                        & Understand-Plan                       & success rate       \\
                                                                  & GPT4AIGChip \cite{fu_gpt4aigchip_2023}      & PRT                            & Generator                      & HLS prompts                            & Accelerator RTL                       & Demo-Aug. Prompt                      & PPA/function       \\
                                                                  & VerilogCoder \cite{ho_verilogcoder_2025}    & SFT                            & Agent                          & NL spec/prompts                        & RTL                                   & Semantic Repair                       & pass@k, compile    \\
                                                                  & OpenLLM-RTL \cite{liu_openllm-rtl_2025}     & DATA                           & Benchmark                      & RTL tasks/specs                        & public benchmarks                     & dataset + benchmark                   & suite pass@k       \\
                                                                  & VeriGen \cite{thakur_verigen_2024}          & SFT                            & Generator                      & partial RTL                            & completed RTL                         & fine-tuned LLM                        & pass@k, tests      \\
                                                                  & AutoChip \cite{autochip}                    & PRT                            & Agent                          & NL spec + feedback                     & verified RTL                          & feedback + tools                      & task success       \\
                                                                  & FVDebug \cite{fvdebug}                      & PRT                            & Agent                          & RTL + cex trace                        & fix/patch                             & LLM + FV tool                         & bug-fix rate        \\
                                                                  & RTLLM \cite{lu_rtllm_2024}                  & EVAL                           & Critic                         & NL spec                                & functional RTL                        & benchmark suite                       & functional pass    \\
                                                                  & VerilogEval \cite{pinckney_revisiting_2025} & EVAL                           & Critic                         & generated RTL                          & scores                                & benchmark suite                       & pass@k              \\
        \midrule
        \multirow{8}{*}{\rotatebox[origin=c]{90}{Backend}}
                                                                  & AlphaChip \cite{mirhoseini_graph_2021}      & FDN                            & Agent                          & netlist + constr.                      & placement                             & RL + pre-training                     & WL/congestion      \\
                                                                  & CodeNotCanvas \cite{yang_code_2026}         & MAS                            & Agent                          & constraints                            & Tcl scripts                           & multi-agent nego.                     & PPA/validity       \\
                                                                  & DiffusionPlace \cite{lee_chip_2025}         & FDN                            & Generator                      & netlist + constr.                      & placement                             & diffusion model                       & cong., timing, WL  \\
                                                                  & TransPlace \cite{hou_transplace_2025}       & FDN                            & Generator                      & netlist + constr.                      & placement                             & GNN + transfer                        & cong., timing, WL  \\
                                                                  & ORFS-Agent \cite{ghose_orfs-agent_2025}     & MAS                            & Agent                          & netlist + constr.                      & optimized soln.                       & agent + solver                        & QoR / runtime      \\
                                                                  & ChatEDA \cite{he_chateda_2024}              & PRT                            & Planner                        & spec + tools                           & reports / GDSII                       & multi-agent orches.                   & WNS/TNS, Pwr        \\
                                                                  & REvolution \cite{min_revolution_2025}       & MAS                            & Agent                          & netlist                                & optimized PPA                         & evolutionary agent                    & PPA Gain           \\
                                                                  & Chip-Chat \cite{blocklove_chip-chat_2023}   & PRT                            & Agent                          & NL conversation                        & Verilog modules                       & conversational assistant              & task success       \\
                                                                  & ArchPower \cite{zhang2025archpower}       & DATA                           & Critic                         & arch. config                           & power est.                            & dataset + proxy                       & accuracy            \\
        \midrule
        \multirow{8}{*}{\rotatebox[origin=c]{90}{Found. \& Sec}}
                                                                  & ChipNeMo \cite{liu_chipnemo_2023}           & FDN                            & Foundation                     & domain corpus                          & found. model                          & DAPT + Tokenizer                      & domain adapt.      \\
                                                                  & SynC-LLM \cite{liu_syncllm_2025}            & DATA                           & Generator                      & (generative)                           & synthetic Verilog                     & hierarchical gen.                     & scale/quality       \\
                                                                  & GenEDA \cite{fang_geneda_iccad_2025}              & DATA                           & Benchmark                      & netlist + text                         & aligned embed.                        & cross-modal align                     & retrieval/QA       \\
                                                                  & AnalogSeeker \cite{chen_analogseeker_2025}  & FDN                            & Foundation                     & analog spec                            & circuit topology                      & knowledge distill.                    & QA accuracy        \\
                                                                  & SoCureLLM \cite{tarek_socurellm_2025}       & MAS                            & Agent                          & SoC design                             & security policy                       & red-teaming agent                     & detection rate     \\
                                                                  & TrojanStego \cite{trojanstego}              & SFT                            & Generator                      & RTL syntax                             & leaked bits                           & steganography                         & bit error rate     \\
                                                                  & DRC-Coder \cite{chang_drc-coder_2025}       & SFT                            & Agent                          & rules + layouts                        & DRC checker code                      & multi-agent VLM                       & F1, time/cost      \\
                                                                  & ChiPBench \cite{wang_benchmarking_2024}     & EVAL                           & Critic                         & design + toolchain                     & PPA outcome                           & end-to-end bench.                     & PPA/QoR/cost       \\
        \bottomrule
    \end{tabular}
    }
\end{table*}

\vspace{0.4cm}
\section{Foundations: The Agentic Cognitive Stack}
\label{sec:foundations}

\begin{figure}[t]
    \centering
    \includegraphics[width=0.99\linewidth]{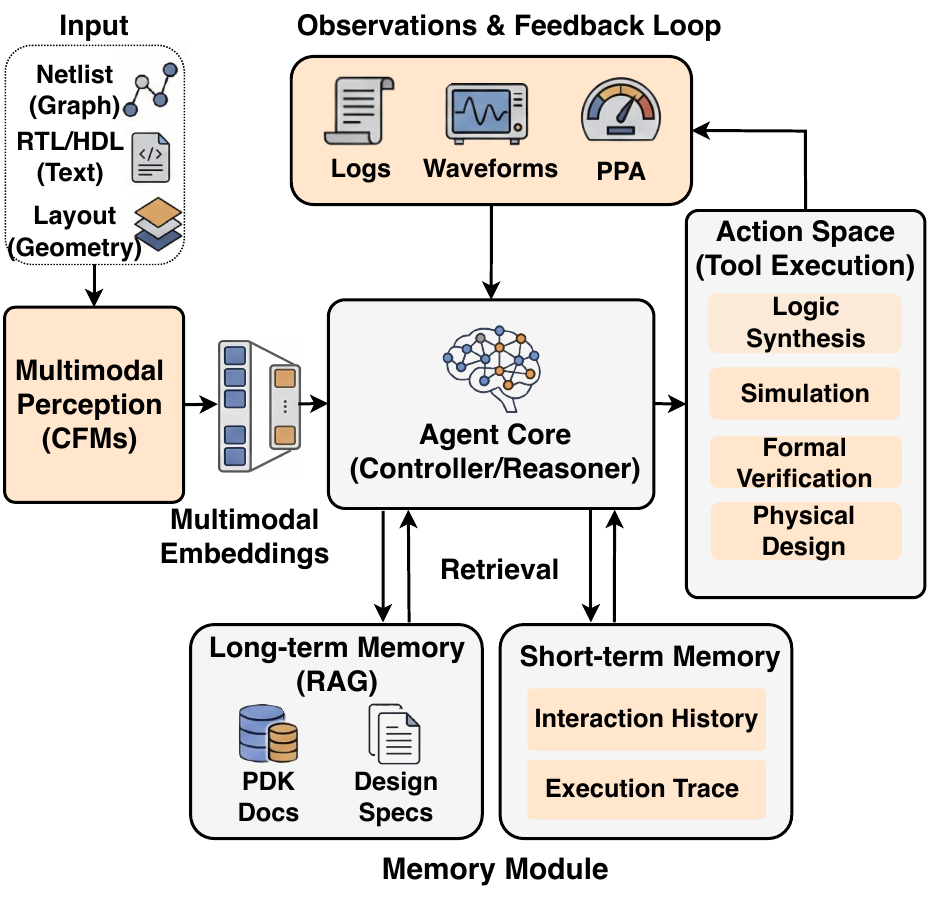}
    \vspace{-0.2cm}
    \caption{\textbf{The Three-Layer Cognitive Stack of Agentic EDA.} We decompose the autonomous design process into: (1) \textbf{Perception}, mapping heterogeneous formats (netlists, logs, layouts) into a unified multimodal latent space; (2) \textbf{Cognition}, the neuro-symbolic core where \textit{Reasoning} and \textit{Planning} occur, supported by \textbf{Hierarchical Memory} (RAG/Trace) to maintain long-horizon context; and (3) \textbf{Action}, which grounds plans into executable scripts (Tcl/Python). Crucially, the system operates as a \textbf{Closed Loop}, using tool feedback (Logs/Waveforms/PPA) to iteratively correct errors and mitigate hallucination.}
    \label{fig:agentic_stack}
\end{figure} 

To analyze the transition to engineer-centric autonomy, we introduce a unifying framework comprising a {Cognitive Stack} and a {Methodological Taxonomy}.

\paragraph{2.1 The Cognitive Stack (Architecture).}
The transition from static LLMs to dynamic Agents implies a specialized architecture. We structure our analysis around a three-layer {Cognitive Stack} (Perception-Cognition-Action) that enables the "Perceive-Think-Act" loop essential for autonomy (Fig. \ref{fig:agentic_stack}).
First, \textbf{Perception (Multimodal Representation)} transforms heterogeneous data—Netlists (Graph), HDL (Text), Layouts (Geometry)—into unified \textit{multimodal embeddings}. Models like \textit{CircuitFusion} \cite{fang_circuitfusion_2025} and \textit{GenEDA} \cite{fang_geneda_iccad_2025} use contrastive learning to align code with graphs, overcoming the \textit{Granularity Mismatch} between hierarchical HDL and flattened netlists, while \textit{DeepGate4} \cite{zheng_deepgate4_2025} addresses scalability via sub-linear memory graph transformers. Addressing \textit{Data Scarcity}, recent works employ \textit{Synthetic Data Generation}: techniques like \textit{instruction back-translation} 
and \textit{bug injection} (e.g., \textit{CraftRTL} \cite{liu_craftrtl_2025}) create massive synthetic corpora to bootstrap agents, overcoming the lack of proprietary industrial-grade IP and scripts.

Second, \textbf{Cognition (Reasoning \& Planning)} acts as the \textit{Neuro-Symbolic Bridge}, resolving the fundamental tension between probabilistic LLMs and deterministic physics. Agents do not directly "calculate" timing or power; rather, they function as \textit{Heuristic Search Engines} that hypothesize solutions, while traditional solvers (SAT/SMT, SPICE) act as \textit{Ground Truth Verifiers}. This layer handles \textit{Hierarchical Planning}, where MAS evolve into \textit{Cooperative Games}: a ``Manager'' arbitrates between specialized agents to resolve deadlock, while a ``Critic'' uses formal verification to prune invalid branches. It also manages \textit{Long-Horizon Consistency} to address ``Design Intent Drift'' in multi-month flows via \textit{Hierarchical Memory Systems} or \textit{Constraint Databases}, and employs \textit{Domain-Specific RAG} to retrieve not just documentation but structured design rules (DRC) and historical error logs to guide reasoning.

Third, \textbf{Action (Tool Execution)} converts reasoning into physical changes. Unlike LLMs that output text, Agents perform NL2SL translation to control EDA tools, which we view as \textit{external deterministic solvers} \cite{cheng_empowering_2025}. This layer ensures \textit{Grounding} and \textit{Safety} by verifying that generated actions adhere to semiconductor physics constraints and executing scripts in sandboxed environments with checkpoint-based rollback to recover from destructive hallucinations.

\begin{figure}[t]
    \centering
    \includegraphics[width=0.99\linewidth]{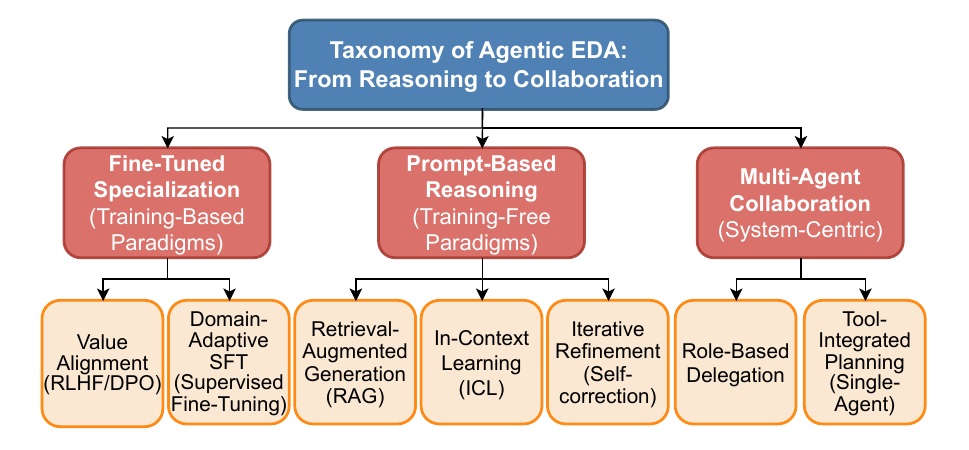}
    \vspace{-0.2cm}
    \caption{Methodological Taxonomy of Agentic EDA Paradigms. We categorize approaches into three tiers of increasing autonomy: (1) \textbf{Prompt-Based Reasoning} (Training-Free), leveraging general-purpose LLMs via ICL and RAG for flexible but superficial adaptation; (2) \textbf{Fine-Tuned Specialization} (Training-Centric), internalizing domain physics via SFT/RLHF for high performance but high cost; and (3) \textbf{Multi-Agent Collaboration} (System-Centric), solving complex long-horizon tasks via multi-agent collaboration and iterative tool use.}
    \label{fig:tax}
\end{figure}

\paragraph{2.2 Methodological Taxonomy (Paradigm).}
As illustrated in Figure \ref{fig:tax}, we categorize existing systems into three tiers based on their implementation complexity.
\textit{Prompt-Based Reasoning (In-Context Learning)} involves training-free agents (e.g., \textit{ChatEDA} \cite{he_chateda_2024}) that leverage general-purpose LLMs via ReAct prompting to refine scripts based on tool feedback.
\textit{Specialized Models (SFT/RLHF/Pre-training)} internalize domain knowledge via Supervised Fine-Tuning. \textit{AlphaChip} \cite{goldie_chip_2024} exemplifies this by acquiring ``layout intuition'' through massive pre-training on chip placements.
\textit{Multi-Agent Collaboration (System-Centric)} represents the most advanced tier, involving Multi-Agent Systems (MAS) that align intent via complex planning. \textit{ArchPower} \cite{zhang2025archpower} demonstrates this by grounding high-level architectural decisions in detailed power modeling through collaborative agent interactions.

\paragraph{2.3 Resolving the "Probabilistic vs. Deterministic" Paradox.}
A critical critique of Agentic EDA is the risk of using probabilistic models for zero-tolerance engineering ("using dice to guide a scalpel"). We argue that this view fundamentally misinterprets the agent's role.
\textit{The Neuro-Symbolic Handshake}: The agent provides \textit{Exploration} (navigating the non-convex search space), while the EDA tool provides \textit{Exploitation} and \textit{Verification} (enforcing physics). The LLM never "guesses" the final sign-off; it generates \textit{candidate scripts} which must pass the deterministic checks of the compiler, timer, and DRC engine. If a check fails, the error log serves as a \textit{Gradient Signal} for the next iteration. Thus, the final output is not probabilistic; it is the result of a probabilistic search grounded in a deterministic verifier \cite{xu_revolution_2025}.
\textit{From Guessing to Reasoning}: Current research aims to shift from \textbf{Type I (Reactionary Retry)}—blindly fixing errors based on logs \cite{blocklove_chip-chat_2023,autochip}—to \textbf{Type II (Planner)}, where agents employ \textit{Lookahead} mechanisms (e.g., Tree of Thoughts) to simulate outcomes \textit{before} execution \cite{cheng_empowering_2025}.

\section{Frontend: RTL Implementation \& Verification Loops}

\begin{figure}[t]
    \centering
    \includegraphics[width=0.99\linewidth]{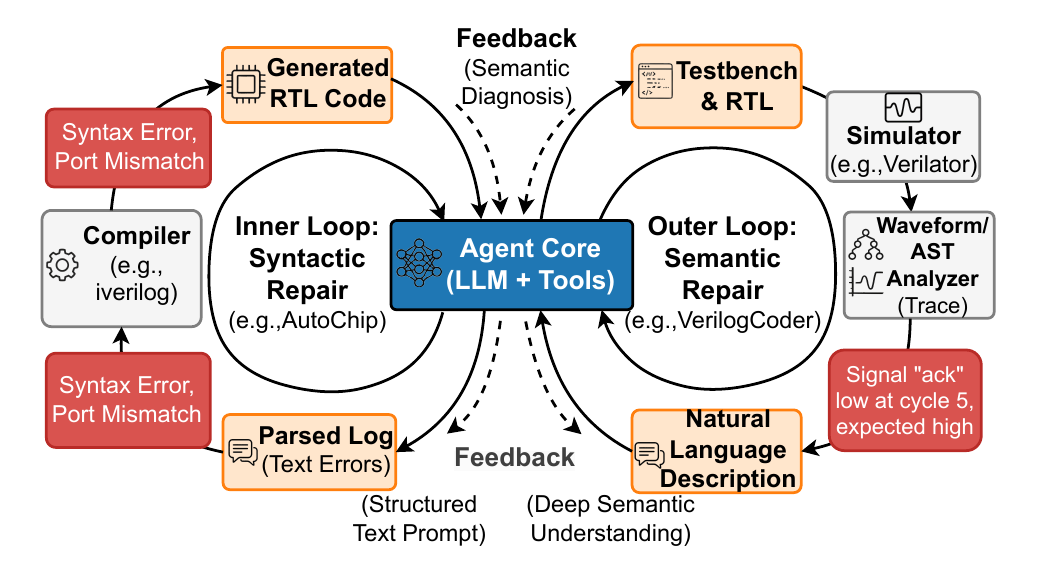} 
    \caption{Dual-Loop Agentic RTL Repair Framework. To balance verification cost and fidelity, the system employs a hierarchical feedback mechanism: a fast \textbf{Inner Loop} (Syntactic Repair) using compiler logs to quickly filter syntax errors, and a slow, high-fidelity \textbf{Outer Loop} (Semantic Repair) leveraging simulation traces and formal verification to correct functional logic bugs. This coarse-to-fine strategy significantly reduces the computational overhead of agentic iterations.}
    \label{fig:repair_loop}
\end{figure}

Frontend design—the translation of architectural intent into functional RTL and its subsequent verification—represents the most mature domain for Agentic EDA. As shown in Figure \ref{fig:repair_loop}, research has evolved from simple ``one-shot generation'' \cite{thakur_verigen_2024} to robust agentic workflows. While models like GPT-4o achieve $\sim$63\% pass rates on \textit{VerilogEval} \cite{pinckney_revisiting_2025}, the complexity of industrial design necessitates a shift from probabilistic token prediction to rigorous engineering loops. We categorize these advancements by their role in the standard IC lifecycle: {RTL Implementation} (Construction) and {Autonomous Verification} (Assurance).

\paragraph{3.1 RTL Implementation (Code Generation \& Iterative Repair).}

RTL implementation is not merely translation but a \textit{Correct-by-Construction} constraint satisfaction problem. Agents address this via an engineering feedback loop.
First, \textit{Spec-to-RTL Translation} tackles the ambiguity of natural language specifications. \textit{Spec2RTL-Agent} \cite{yu_spec2rtl_2025} employs an ``Understanding-Planning-Coding'' workflow, treating coding as a multi-step reasoning task rather than simple completion.
Second, the \textit{Dual-Loop Repair Mechanism} is the core engine for correctness (Fig. \ref{fig:repair_loop}). To mitigate hallucination, agents utilize a hierarchical feedback system: a fast \textit{Syntactic Loop} uses compiler logs to filter syntax errors (e.g., \textit{AutoChip} \cite{autochip}), while a slow, high-fidelity \textit{Semantic Loop} traces functional causality via simulation waveforms or ASTs (e.g., \textit{VerilogCoder} \cite{ho_verilogcoder_2025}).
Third, \textit{Instruction Alignment} aligns probabilistic outputs with deterministic protocols. Techniques like Structured Instruction Chain-of-Thought (SI-CoT) \cite{haven} enforce a ``think-before-code'' policy, significantly reducing syntax violations by grounding generation in explicit design rules.

\paragraph{3.2 Autonomous Verification (The DV Flow).}

Verification consumes over 60\% of the design cycle. We restructure recent AI advances to map directly onto the standard Design Verification (DV) lifecycle, highlighting how agents augment each stage with reasoning capabilities.
\begin{enumerate}[leftmargin=*, label=\arabic*]
    \item \textit{Test Planning \& Stimulus Generation}: The first step is defining \textit{what} to verify. Agents like \textit{Saarthi} \cite{saarthi} autonomously decompose specifications into comprehensive verification plans. For stimulus, \textit{PRO-V-R1} \cite{pro_v_r1} orchestrates agents to generate Python-based test vectors (e.g., UVM sequences), effectively exploring corner cases that rule-based random testing might miss.
    \item \textit{Assertion \& Formal Grounding}: To verify correctness, agents must translate intent into formal logic. \textit{SANGAM} \cite{gupta_sangam_2025} employs Monte Carlo Tree Search (MCTS) to navigate the space of SystemVerilog Assertions (SVA), transforming ``Text Generation'' into a rigorous ``Logic Search.'' Similarly, \textit{VeriMaAS} \cite{pamnani_ai-driven_2025} integrates Formal Verification (FV) solvers as a \textit{Grounding Mechanism}, ensuring that generated properties are mathematically consistent.
    \item \textit{Automated Debugging (Root Cause Analysis)}: When tests fail, agents perform reasoning on failure logs. \textit{FVDebug} \cite{fvdebug} leverages Causal Graphs to perform reverse-reasoning on failure traces, moving beyond simple pattern matching to identify the specific RTL lines responsible for logic violations.
\end{enumerate}

\paragraph{3.3 Benchmarking (Beyond Unit Tests).}

A critical barrier to adoption is the \textit{Unit-Test Fallacy}. The field relies heavily on \textit{VerilogEval} \cite{liu_verilogeval_2023}, which tests module-level syntax. However, high pass rates on such ``toy benchmarks'' can be misleading—akin to judging an aircraft engineer by their ability to fold paper planes. We advocate a pivot toward \textit{System-Level Benchmarks} (e.g., \textit{RTLLM} \cite{lu_rtllm_2024}, \textit{ChipNeMo} \cite{liu_chipnemo_2023}) that evaluate interface protocols, clock domain crossings, and multi-module integration, where the complexity lies in managing constraints and system-level dependencies.
 
\section{Backend: Physical Design (Algorithms \& Tool Orchestration)}
\label{sec:backend}

Physical design, particularly placement and routing (P\&R), represents the most computationally intensive phase. As summarized in Table \ref{tab:backend_comparison}, this domain is witnessing a dichotomy between \textit{Algorithm-Centric} solvers and \textit{Agent-Centric} workflows. We categorize innovations into two engineering layers: \textit{Neural-Augmented Algorithms} (improving the internal engines) and \textit{Autonomous Tool Orchestration} (automating the external design flow).

\begin{table*}[t]
    \centering
    \caption{Comparative analysis of backend optimization paradigms: Solvers vs. Orchestrators. We contrast \textit{Algorithm-Centric} methods (RL, GNN, Generative) that optimize specific point metrics (e.g., wirelength) against \textit{Agent-Centric} workflows (LLM Agents) that orchestrate the entire design flow. While solvers offer superior inference speed and precision for defined tasks, agents provide unique capabilities in \textbf{Tool Interoperability} and \textbf{End-to-End Orchestration}, albeit with current challenges in convergence stability.}
    \vspace{-0.3cm}
    \label{tab:backend_comparison}
    \small
    \setlength{\tabcolsep}{4pt}
    \renewcommand{\arraystretch}{1.2}
    \begin{tabular}{@{}p{1.6cm}p{2.4cm}p{5.2cm}ccccp{3.2cm}@{}}
        \toprule
        \textbf{Category} & \textbf{Mechanism} & \textbf{Representative Works} & \textbf{Gen.} & \textbf{Train} & \textbf{Infer.} & \textbf{PPA} & \textbf{Primary Limitation} \\ \midrule
        \textbf{RL} & Sequential MDP & \textit{AlphaChip} \cite{mirhoseini_graph_2021}, \textit{MaskPlace} \cite{lai_maskplace_2022} & Low & High & Slow & High & Poor Transferability \\
        \textbf{Generative} & Cond. Denoising & \textit{DiffusionPlace} \cite{lee_chip_2025} & High & High & Fast & Med & Lack of Precision \\
        \textbf{GNN} & Graph Embedding & \textit{TransPlace} \cite{hou_transplace_2025}, \textit{DeepGate4} \cite{zheng_deepgate4_2025} & High & Med & Fast & High & Graph Scalability \\
        \textbf{Agentic} & Tool Orchestration & \textit{CodeNotCanvas} \cite{yang_code_2026}, \textit{ChatEDA} \cite{he_chateda_2024} & High & Low & Slow & High & Stability/Convergence \\
        \bottomrule
    \end{tabular}
    \vspace{-0.6em}
\end{table*}

\paragraph{4.1: Neural-Augmented Algorithms (Solvers).}

This level focuses on replacing heuristic components of EDA tools with learned models. A central engineering challenge is the choice of \textit{Latent Representation}. Early ``Canvas'' approaches treated layout as image generation \cite{lai_maskplace_2022}, applying CNNs to predict macro locations. However, pixel-based representations are fundamentally misaligned with EDA constraints for two reasons: first, \textit{discretization error}, where mapping continuous nanometer coordinates to discrete grid pixels introduces quantization noise that invariably violates Design Rule Checks (DRC); and second, \textit{non-local dependencies}, where pixel correlations are local, whereas in a chip, a critical timing path may connect two spatially distant gates.
Consequently, algorithms are evolving from \textit{Sequential RL to Generative Diffusion}. Reinforcement Learning (RL), epitomized by Google's \textit{AlphaChip} \cite{mirhoseini_graph_2021}, dominates industrial macro placement and has achieved successful deployment in TPU production. However, it faces two persistent challenges: \textit{Sequential overfitting}, where formulating placement as a Sequential MDP often leads to memorizing specific netlist structures rather than learning generalizable policies; and \textit{Non-differentiable tools}, as commercial EDA engines are black boxes, forcing reliance on sample-inefficient optimization. In contrast, Diffusion models \cite{lee_chip_2025} redefine placement as a Conditional Denoising task $p_\theta(x|c)$. By iteratively refining random coordinates guided by connectivity gradients, this paradigm not only improves zero-shot generalization across unseen netlists but also significantly reduces inference overhead compared to the repetitive sampling required by sequential policies.

\begin{table*}[t]
    \centering
    \caption{Taxonomy of emerging security landscapes in Agentic EDA. The introduction of autonomous agents expands the attack surface into two dimensions: \textbf{Model-Centric Risks} (e.g., Trojans embedded in the agent's weights or prompts) and \textbf{Infrastructure-Centric Risks} (e.g., privacy leakage via cloud APIs). The table outlines corresponding defense mechanisms, highlighting the critical need for \textbf{Trustworthiness Metrics} to distinguish between stochastic hallucinations and malicious adversarial attacks.}
    \label{tab:security_taxonomy}
    \vspace{-0.2cm}
    \footnotesize
    \renewcommand{\arraystretch}{1.0}
    \begin{tabular}{@{}p{2.0cm}p{1.6cm}p{4.0cm}p{4.0cm}p{4.4cm}@{}}
        \toprule
        \textbf{Domain} & \textbf{Threat} & \textbf{Mechanism} & \textbf{Defense Strategy} & \textbf{Representative Works} \\ \midrule
        \multirow{6}{2.0cm}{\centering\textbf{Model Integrity}} 
        & \textbf{Hardware Trojans} & Malicious logic insertion via LLM suggestions; Information leakage through code structure & Intent Analysis, Anomaly Detection, Code Sanitization & TrojanStego \cite{trojanstego}, TrojanWhisper \cite{trojanwhisper}, LATENT \cite{latent} \\ \cmidrule(l){2-5}
        & \textbf{Hallucinations} & Protocol violations, invalid syntax generation, semantic misalignment & Structured Chain-of-Thought (CoT), Formal Verification (FV) Feedback & HaVen \cite{haven}, VeriMaAS \cite{pamnani_ai-driven_2025} \\ \midrule
        \multirow{5}{2.0cm}{\centering\textbf{Infrastructure}} 
        & \textbf{Compromised Artifacts} & Flawed/Malicious DRC rule decks; Tool-facing script injection & Auditable Generation, Rule Reasoning, Deterministic Checkers & DRC-Coder \cite{chang_drc-coder_2025}, SoCureLLM \cite{tarek_socurellm_2025} \\ \cmidrule(l){2-5}
        & \textbf{Data Privacy} & Training data leakage, IP exposure in cloud-based inference & Federated Learning, Local Deployment, Differential Privacy & CircuitNet \cite{chai2023circuitnet} \\ \bottomrule
    \end{tabular}
\end{table*}

\paragraph{4.2: Autonomous Tool Orchestration (The "Flow").}

Unlike solvers that optimize a fixed objective, Agentic AI acts as a hierarchical \textit{Orchestrator} to close the design loop by managing commercial tools.
First, \textit{Script Generation \& Interface} treats ``Code-as-Action''. By generating executable Tcl constraints \cite{yang_code_2026}, agents leverage the commercial toolchain as a reliable ``Decoder'' that transforms symbolic plans into legal layouts. The key innovation here is \textit{Resilience}: agents parse tool logs (`stderr`) and self-correct scripts using RAG, transforming brittle automation into robust engineering workflows.
Second, \textit{Cross-Stage Closure} represents the ultimate promise of automating ECOs. Uniquely, agents can parse backend reports (e.g., negative slack or congestion maps) and autonomously modify RTL (e.g., inserting pipeline stages). This shifts from "Blind Generation" to \textit{physical-aware code refinement}. \textit{ChatEDA} \cite{he_chateda_2024} exemplifies this by decomposing high-level requests into tool scripts with a 98.3\% success rate. Similarly, \textit{Chip-Chat} \cite{blocklove_chip-chat_2023} demonstrates how conversational agents can iteratively refine design intent. Beyond correctness, \textit{REvolution} \cite{min_revolution_2025} uses evolutionary agents to continuously optimize PPA, achieving a 24.5\% power reduction. Critical to this loop are fast proxies like \textit{ArchPower} \cite{zhang2025archpower}, which estimate power instantly to guide iterative \textit{RTL-to-GDSII-to-RTL} refinement.
Going beyond PPA, agents are beginning to tackle \textit{system-level co-design}, jointly optimizing hardware accelerators and their compiler stacks (e.g., mapping tensor operators to custom ISA) to maximize end-to-end application performance \cite{fu_gpt4aigchip_2023}. This capability to span the full stack—from software kernels to transistor sizing—demonstrates the true potential of \textit{agentic autonomy} over isolated point tools.

\textbf{Emerging Frontier: Analog/Mixed-Signal (AMS) Design.}
While the aforementioned frameworks focus on digital logic, the agentic paradigm is also transforming the analog domain. Unlike digital's discrete abstraction, AMS design requires tuning continuous parameters (e.g., transistor widths) under strict physical constraints. \textit{AnalogSeeker} \cite{chen_analogseeker_2025} employs knowledge distillation to transfer circuit intuition to agents, while \textit{Gao et al.} \cite{gao2024post} demonstrate that Agent-guided Bayesian Optimization outperforms human experts in operational amplifier sizing by efficiently navigating the non-convex PPA landscape.
  
\section{Trustworthiness \& Standardization}
\label{sec:security}

The transition from predictive tools to autonomous agents introduces a fundamental \textit{Trust Deficit}. Unlike traditional EDA algorithms (e.g., A* search), which are deterministic and mathematically bounded, LLM-based agents are probabilistic and opaque. As detailed in Table \ref{tab:security_taxonomy}, this shift expands the risk surface from simple bugs to complex trust issues. Operationalizing Agentic EDA thus requires a transition from ``Human-in-the-Loop'' validation to rigorous {Trust Engineering}, ensuring that generative creativity is strictly bounded by physical constraints and explainable logic.

\paragraph{5.1 The Trust Deficit: Hallucinations vs. Trojans.}
Trustworthiness in Agentic EDA is compromised by both stochastic errors and malicious intent. First, \textit{Hallucinations (Stochastic Risks)} occur when agents generate syntactically correct but functionally flawed designs (e.g., connecting a clock signal to a reset pin). Unlike natural language where ambiguity is acceptable, hardware requires zero-tolerance precision. Second, \textit{Trojans (Adversarial Risks)} emerge from the opacity of large models. \textit{TrojanStego} \cite{trojanstego} demonstrates how attackers can embed malicious logic (Hardware Trojans) into generated RTL via subtle code structure manipulations, while \textit{TrojanWhisper} \cite{trojanwhisper} highlights the risk of ``Supply Chain Attacks'' where poisoned training data leads agents to insert backdoors. Defense requires a multi-layered approach: \textit{Intent Analysis} to audit the agent's reasoning chain (CoT) before execution, and \textit{Red Teaming Agents} like \textit{SoCureLLM} \cite{tarek_socurellm_2025} that autonomously probe designs for vulnerabilities.

\paragraph{5.2 Verification \& Explainability (The "Why" Problem).}
For IC engineers to trust an agent, correctness is insufficient; \textit{Explainability} is mandatory. A "black-box" optimization that magically fixes a timing violation without justification is professionally unacceptable. We propose framing Formal Verification not just as a checker, but as an \textit{Explainability Mechanism}.
\textit{Traceability}: Agents must provide a \textit{Reasoning Trace} (Chain-of-Thought) linking the high-level intent ("fix setup violation") to the low-level action ("insert buffer at net X"), allowing engineers to audit the decision logic.
\textit{Proof-Carrying Code}: Borrowing from software security, future agents should output not just RTL, but accompanying \textit{Formal Properties} (SVA) that prove the generated code meets the spec. This shifts the trust model from trusting the \textit{Model} (which is fallible) to trusting the \textit{Proof} (which is mechanically verifiable) \cite{haven}.

\paragraph{5.3 Standardization \& Benchmarks (The "How" Problem).}
The fragmentation of tool interfaces is a primary barrier to building trustworthy agents. Currently, agents interact with EDA tools via fragile, tool-specific text parsing.
\textit{Open Agentic EDA Standard}: To enable robust autonomy, the field needs a standardized interface protocol (analogous to the Language Server Protocol in SWE) that exposes EDA tool states (e.g., timing graphs, congestion maps) as structured, machine-readable APIs rather than raw log text \cite{qiu_aieda-20_2025}.
\textit{System-Level Benchmarking}: Trust is also hindered by the \textit{proxy metric fallacy}. High pass rates on isolated module generation (\textit{VerilogEval}) do not correlate with success in complex flows. We advocate for system-level benchmarks like \textit{ChipNeMo} \cite{liu_chipnemo_2023} and \textit{ChiPBench} \cite{wang_benchmarking_2024} that evaluate end-to-end PPA (Power, Performance, Area) outcomes. Only by demonstrating reproducibility on industrial-grade constraints can Agentic EDA cross the chasm from academic novelty to engineering adoption.
 
\section{Challenge and Future Directions}

Generative AI is reshaping Electronic Design Automation (EDA) from tool-driven automation toward higher levels of workflow autonomy. However, the transition remains constrained by reliability under a zero-tolerance error regime, limited and proprietary training data, and the opacity of legacy toolchains—sustaining the debate on whether the current wave is a ``revolution or hype'' \cite{xu_revolution_2025}. These constraints indicate that progress cannot rely on scaling alone.

\textbf{The Grand Challenge: The Sim-to-Silicon Gap.}
A critical criticism of the current Agentic EDA landscape is the lack of ``Ground Truth'' validation via actual tape-outs. It is crucial to distinguish between \textit{Silicon-Proven Solvers} and \textit{Agentic Prototypes}. While RL-based systems (e.g., AlphaChip \cite{mirhoseini_graph_2021}) have successfully achieved industrial tape-out, no LLM-driven autonomous agent has yet demonstrated a complete, human-free tape-out of an industrial chip. Most academic agents optimize for proxy metrics (e.g., placement wirelength) or use open-source PDKs (e.g., Sky130) that do not reflect the complex physical effects of advanced nodes (e.g., TSMC N3). Critics argue that agents trained on such ``toy data'' cannot generalize to industrial scenarios.
We argue that while parameter-level generalization is limited, \textit{methodological transferability} remains valid: the \textit{logic} of debugging a hold violation (e.g., inserting buffers) is invariant across process nodes, even if the specific buffer drive strengths differ. To bridge the data gap, the community must prioritize \textit{industrial-grade benchmarks}, moving beyond obfuscated benchmarks to synthetic datasets that replicate advanced-node constraints (e.g., multi-patterning, electromigration) via procedural generation. Simultaneously, \textit{domain adaptation techniques} must be developed to transfer reasoning patterns from data-rich open domains to data-poor proprietary domains without leaking IP (e.g., via Federated Learning).

\textbf{Future Direction I: Technical Grounding (Reliability \& Interface).}
Deploying autonomy requires solving technical alignment. First, \textit{Constraint Grounding} addresses the conflict between probabilistic LLMs and zero-tolerance hardware. Future work must achieve \textit{Neuro-Symbolic Alignment}, where generative outputs are mathematically bounded by formal verification constraints \cite{haven}. Second, \textit{Tool Grounding (Standardized Interfaces)} is essential as agents currently interact with tools via fragile text logs. The future lies in \textit{Semantic EDA Interfaces}—APIs exposing internal tool states (e.g., timing graphs) as structured embeddings. This necessitates an \textit{Open Agentic EDA Standard} (analogous to LSP) to unify how agents query tool states \cite{qiu_aieda-20_2025}.

\textbf{Future Direction II: Operational Grounding (Deployment \& Economics).}
\textit{Economic Grounding (The Cost of Autonomy)} addresses skepticism regarding the high inference cost of agents (e.g., \$30 per debugging session via GPT-4o) compared to human labor. However, this comparison overlooks the \textit{Time-to-Market (TTM)} premium. In an industry where missing a tape-out window costs millions, the value of an agent lies not in being cheaper than an intern, but in being \textit{faster} than a burnt-out engineer. By offloading the "grind" of iterative Tcl debugging to agents (which run 24/7), design teams can reduce the critical path schedule. Future benchmarks must therefore report \textit{Wall-clock Efficiency} (Time-to-Result) alongside token costs to capture this trade-off. Furthermore, \textit{Failure Mode Taxonomy} is required for robustness, characterizing agent failures such as \textit{Deadlooping} (oscillation without convergence), \textit{Context Loss} (forgetting global constraints), and \textit{Tool Hallucination}.

\textbf{Visionary Outlook: Recursive Self-Improvement.}
Beyond productivity, Agentic EDA points toward a speculative but transformative direction: \textit{AI Designing AI Hardware}. As agents become capable of optimizing novel architectures, they may assist in designing the next generation of chips that train stronger agents. While this \textit{Recursive Self-Improvement} loop remains a distant vision, it suggests that \textit{Constrained Neuro-Symbolic Planning} could provide a blueprint for other physical engineering domains.

\textbf{Pragmatic Roadmap.}
\textit{Near-term (1-2 years)} sees L3 agents as ``strong copilots'' for script generation (Tcl) with human oversight. \textit{Mid-term (3-5 years)} anticipates autonomous block-level flows, provided that Sim-to-Real benchmarks bridge gaps. \textit{Long-term (5+ years)} envisions full-chip agentic architecture exploration requiring certified guarantees.

\clearpage

\bibliographystyle{named}
{\footnotesize
\bibliography{common.bib,intro.bib,foundations.bib,frontend.bib,backend.bib,future.bib,unused.bib}  
}
\end{document}